\documentclass[aps,prc,preprint,amsfonts,amsmath,superscriptaddress,floatfix,nofootinbib]{revtex4-2}
\usepackage{graphicx,epsfig,color,amsmath, booktabs}
\usepackage{soul}
\usepackage{physics}
\usepackage{mathtools}
\usepackage{mathrsfs}
\DeclareMathAlphabet{\mathpzc}{OT1}{pzc}{m}{it}
\usepackage{booktabs}

\newcommand{\beqy}{\begin{eqnarray}}
	\newcommand{\eeqy}{\end{eqnarray}}

\newcommand{\h}{h_q}
\newcommand{\D}{\Delta_q}

\newcommand{\mubar}{\bar{\mu}_q}
\newcommand{\Vbar}{\bar{\mathbb{V}}_q}
\newcommand{\Tbar}{\bar{T}_q}
\newcommand{\Dbar}{\bar{\Delta}_q}
\newcommand{\Lbar}{\bar{\varepsilon}_{\Lambda}}
\newcommand{\Ex}{\mathbb{E}_x^{(q)}}

\newcommand{\Fermi}{\varepsilon_{Fq}}

\newcommand{\Cutoff}{\varepsilon_{\Lambda}}

\newcommand{\PositiveBound}{\delta_q^{+}}
\newcommand{\NegativeBound}{\delta_q^{-}}
\newcommand{\VLq}{\mathbb{V}_{Lq}}
\newcommand{\Vq}{\mathbb{V}_{q}}
\newcommand{\Vcq}{\mathbb{V}_{cq}^{(0)}}
\newcommand{\Tcq}{T_{cq}^{(0)}}
\newcommand{\Cnorm}{c_N^{(q)}}
\newcommand{\expgamma}{\text{e}^{\gamma}}

\newcommand{\rb}{\pmb{r}}
\newcommand{\rp}{\pmb{r^\prime}}

\newcommand{\sg}{\sigma}
\newcommand{\sgp}{\sigma^\prime}

\newcommand{\f}{f^{(q)}}

\newcommand{\Euler}{\text{e}}
\newcommand{\DoS}{\mathscr{D}_q\left(\mathcal{E},\Delta_q(T,\mathbb{V}_q)\right)}
\newcommand{\QuasipartEnergy}{\mathfrak{E}_{\pmb{k}}^{(q)}}
\newcommand{\DqZero}{\mathcal{D}_q (0)}
\newcommand{\DNq}{\mathscr{D}_N^{(q)}(0)}
\newcommand{\mqOplus}{m_q^{\oplus}}

\begin{document}

	\title{Gapless superfluidity in neutron stars: Thermal properties}
	
	\author{V. Allard}
	\affiliation{Institute of Astronomy and Astrophysics, Universit\'e Libre de Bruxelles, CP 226, Boulevard du Triomphe, B-1050 Brussels, Belgium}
	
	\author{N. Chamel}
	\affiliation{Institute of Astronomy and Astrophysics, Universit\'e Libre de Bruxelles, CP 226, Boulevard du Triomphe, B-1050 Brussels, Belgium}

	\date{\today}

	\begin{abstract}
		The interior of mature neutron stars is expected to contain superfluid neutrons and superconducting protons. The influence of temperature 
		and currents on superfluid properties is studied within the self-consistent time-dependent nuclear energy-density functional theory. We find that 
		this theory predicts the existence of a regime in which nucleons are superfluid (the order parameter remains finite) even though the energy spectrum of quasiparticle excitations 
		exhibits no gap. We show that the disappearance of the gap leads to a specific heat that is not exponentially suppressed at low temperatures 
		as in the BCS regime but can be comparable to that in the normal phase. Introducing some dimensionless effective superfluid velocity, we show 
		that the behavior of the specific heat is essentially universal and we derive general approximate analytical formulas
		for applications to neutron-star cooling simulations. 
	\end{abstract}

	\maketitle

	\section{Introduction}
	
	The densely packed material constituting neutron stars several decades after their formation is thought to become cold 
	enough for the existence of various superfluid and superconducting phases~\cite{chamel2017} (for a review about neutron-star cooling, see, e.g., Ref.~\cite{potekhin2015}): 
	neutron superfluidity with $^1$S$_0$ type pairing in the inner crust and outer core and with $^3$PF$_2$
	pairing in deeper regions, proton superconductivity with $^1$S$_0$ pairing in the outer core, and
	possibly more exotic phases involving hyperons or quarks in the inner core. Predicted by Arkhady Migdal in 1959~\cite{migdal1959}
	and extensively studied theoretically since then~\cite{sedrakian2019}, nuclear superfluidity has found  
	support from radio-timing  observations of pulsar frequency glitches~\cite{antonopoulou2022}, 
	and more recently from the rapid decline of luminosity of the youngest known neutron star in the supernova 
	remnant of Cassiopeia A~\cite{page2011,shternin2011,elshamouty2013,posselt2018,wijngaarden2019,wynnho2021,posselt2022}. 
	
	So far, most microscopic calculations of the nuclear pairing properties have been carried out for superfluids at rest (see, e.g., Ref.~\cite{sedrakian2019}
	for a recent review); a notable exception is the study of Ref.~\cite{alm1996} about the 
	possibility of a $^3$D$_2$ neutron-proton superfluid phase in the inner core of massive neutron stars. However, observed neutron stars are not static but are spinning 
	with periods ranging from milliseconds to seconds, as reported in the Australia Telescope National Facility 
	(ATNF) online catalog\footnote{\url{https://www.atnf.csiro.au/people/pulsar/psrcat/}} (see Ref.~\cite{manchester2005}).
	The neutron and proton superfluids in a neutron star are expected to be rotating at different rates due to 
	their weak coupling with the rest of the star.
	The influence of superflows on the order parameter of the neutron and proton 
	superfluid phases in neutron stars has been previously studied in Refs.~\cite{gusakov2013,leinson2017,leinson2018} 
	using Landau's theory of Fermi liquids. 
	Recently, we have investigated the dynamics of neutron-proton superfluid mixtures within the 
	self-consistent time-dependent nuclear energy-density functional theory. We have obtained exact analytical 
	solutions of the time-dependent Hartree-Fock-Bogoliubov (TDHFB) equations at finite temperatures in homogeneous matter with stationary flows~\cite{ChamelAllard2019,ChamelAllard2020}. Applying this formalism 
	to neutron stars, we have calculated numerically the $^1$S$_0$ neutron and proton pairing gaps 
	in their outer core in the presence of arbitrary currents~\cite{ChamelAllard2021}.

	In this paper, the breakdown of nucleon superfluidity and superconductivity is more closely 
	examined within the same microscopic framework. The critical temperature and critical velocities for the disappearance of nucleon superfluidity and superconductivity are calculated explicitly. 
	We find that the nuclear energy-density functional theory predicts the existence of a regime 
	in which nucleons 
	remain superfluid (the order parameter is finite) even though the energy spectrum of quasiparticle
	excitations exhibits no gap. Such kind of gapless phases have been known for a long time 
	in terrestrial superconductors~\cite{Parmenter1962} and superfluids~\cite{Vollhardt1978} but have 
	not been studied so far in the context of neutron stars. 
	We analyze in details the disappearance of the gap in nuclear superfluids by calculating the density of quasiparticle states. 
	Implications of gapless nuclear superfluidity for the specific heat are investigated. 
	
	The paper is organized as follows. In Section~\ref{sec:theory}, we review the time-dependent nuclear energy-density
	functional theory and its application to homogeneous superfluid mixtures. The properties 
	of nuclear superfluids in neutron stars are presented in Section~\ref{sec:properties}.

	\section{Time-dependent nuclear energy-density functional theory}
	\label{sec:theory}
	
	\subsection{Time-dependent Hartree-Fock-Bogoliubov equations}
	
	The time-dependent density functional theory provides a unified microscopic framework 
	for studying the dynamics of various fermionic systems from atomic nuclei and 
	cold atomic gases to neutron stars  (see, e.g., Ref.~\cite{bulgac2019}), in terms of 
	independent quasiparticle excitations in self-consistent mean fields (see, e.g., Refs.~\cite{blaizot1986,schunck2019}).  Although the nuclear energy-density 
		functional theory is close in spirit to the density functional theory in condensed-matter physics, it is conceptually different because nucleons are self-bound unlike electrons. In nuclear physics, functionals have been 
	traditionally constructed from effective nucleon-nucleon interactions within the mean-field 
	approximation. Although such interactions could in principle be derived from many-body theory~\cite{negele1972} (see also Ref.~\cite{Furnstahl2020} for a recent review of the 
	developments of more microscopically grounded functionals), phenomenological interactions 
	are generally adopted. Correlations are then taken into account in an effective way by introducing density-dependent terms in the interactions and by 
		fitting the parameters to some properties of finite nuclei and infinite 
	nuclear matter.   In the following, we will consider semi-local functionals, such as those obtained from contact interactions of the Skyrme type~\cite{bender03}.
	
	Introducing the set of quantum numbers $k$, 
	the evolution of the two components $\psi_{1k}^{(q)}(\rb,\sg;t)$ and $\psi_{2k}^{(q)}(\rb,\sg;t)$ 
	of the quasiparticle wave function for nucleon species $q$ (here $q=n,p$ for neutrons, protons) 
	at position $\rb$ with spin projection $\sigma=\pm1$ (in units of $\hbar/2$) is governed 
	by the TDHFB equations
	\beqy
	\begin{pmatrix}\h(\rb,t) -\lambda_q & \D(\rb,t) \\ \D(\rb,t)^* & - \h(\rb,t)^*+\lambda_q  \end{pmatrix}\begin{pmatrix} \psi_{1k}^{(q)}(\rb,\sg;t) \\ \psi_{2k}^{(q)}(\rb,\sg;t)\end{pmatrix}=i\hbar \frac{\partial}{\partial t}\begin{pmatrix} \psi_{1k}^{(q)}(\rb,\sg;t) \\ \psi_{2k}^{(q)}(\rb,\sg;t)\end{pmatrix} \, ,
	\label{eq:TDHFB-Russian}
	\eeqy
	where $\lambda_q$ is  the chemical potential. 
	The single-particle Hamiltonian $\h(\rb,t)$ and the pair potential $\D(\rb,t)$ are defined in terms of functional  derivatives  of the total energy $E$ of a matter element of volume $V$ with respect to the  following densities: 
	
	(i) the particle number density at position $\rb$ and time $t$
	\begin{align}\label{eq:Density}
		n_q(\rb,t)=\sum_{\sg=\pm 1} n_q(\rb,\sg;\rb,\sg; t) \, , 
	\end{align}
	
	(ii) the kinetic-energy density (in units of $\hbar^2/2m_q$, with $m_q$ the relevant nucleon mass) at position $\rb$ and time $t$ 
	\begin{align}\label{eq:MomentumDensity}
		\tau_q(\rb,t)=\sum_{\sg=\pm 1}\int\text{d}^3\rp\; \delta (\rb-\rp) \grad\cdot\grad\pmb{'} n_q(\rb,\sg;\rp,\sg; t) \, , 
	\end{align}
	
	(iii) the momentum density (in units of $\hbar$) at position $\rb$ and time $t$
	\begin{align}
		\pmb{j_q}(\rb,t)=-\frac{ i}{2}\sum_{\sg=\pm 1}\int\,{\rm d}^3\rp\,\delta(\rb-\rp) (\pmb{\nabla} -\pmb{\nabla^\prime})n_q(\rb, \sg; \rp, \sg; t)\, ,
	\end{align}
	
	(iv) the pair density\footnote{Also called \emph{anomalous} or \emph{abnormal density}.} at position $\rb$ and time $t$
	\begin{align}
		\widetilde{n}_q(\rb,t)=\sum_{\sg = \pm 1}\widetilde{n}_q (\rb,\sg ; \rb,\sg ; t)\, . 
	\end{align}
	The particle density and pair density matrices~\cite{dobaczewski1984} are defined by the thermal averages
	\begin{equation}\label{eq:ParticleDensityMatrixDefinition}
		n_q(\rb, \sg; \rp, \sgp; t) = <c_q(\rp,\sgp; t)^\dagger c_q(\rb,\sg;t)>\, ,
	\end{equation}
	\begin{equation}\label{eq:PairDensityMatrixDefinition}
		\widetilde{n}_q(\rb, \sg; \rp, \sgp;t) = -\sgp <c_q(\rp,-\sgp;t) c_q(\rb,\sg;t)>\, , 
	\end{equation}
	respectively where $c_q(\rb,\sg;t)^\dagger$ and $c_q(\rb,\sg;t)$ are the creation and destruction operators for particle species of type $q$  at position $\rb$ with spin projection $\sg$ at time $t$. 
	The single-particle Hamiltonian $\h(\rb,t)$ is explicitly given by 
	\beqy
	\label{eq:Hamiltonian}
	\h(\rb,t)=-\pmb{\nabla}\cdot \frac{\hbar^2}{2 m_q^\oplus(\rb,t)}\pmb{\nabla} + U_q(\rb,t)-
	\frac{i}{2}\biggl[\pmb{I_q}(\rb,t)\cdot\pmb{\nabla}+\pmb{\nabla}\cdot\pmb{I_q}(\rb,t)\biggr] \, ,
	\eeqy
	with 
	\beqy
	\label{eq:def-fields}
	\frac{\hbar^2}{2 m_q^\oplus(\rb,t)}=\frac{\delta E}{\delta \tau_q(\rb,t)}, \qquad U_q(\rb,t)=\frac{\delta E}{\delta n_q(\rb,t)},\qquad 
	\pmb{I_q}(\rb,t)=  \frac{\delta E}{\delta \pmb{j_q}(\rb,t)}\, . 
	\eeqy
	The Bogoliubov-de Gennes equations~\cite{degennes} originally developed for inhomogeneous terrestrial superconductors (and later adapted to cold atoms) are recovered after replacing the effective mass $m_q^\oplus(\rb,t)$ by the bare mass $m_q$, and ignoring the potential vector $\pmb{I_q}(\rb,t)$. The pair potential is defined by  
	\beqy
	\label{eq:pair-pot}
	\D(\rb,t) =2\frac{\delta E}{\delta \widetilde{n}_q(\rb,t)^*}\, .
	\eeqy
	Since the energy $E$ is real, it can only depend on the pair density through its square modulus $\vert\widetilde{n}_q(\rb,t)\vert^2$. The pairing potential~\eqref{eq:pair-pot} can thus be written as
	\begin{equation}\label{eq:pair-pot2}
		\D(\rb,t) = 2\frac{\delta E}{\delta \vert\widetilde{n}_q(\rb,t)\vert^2}\widetilde{n}_q(\rb,t) \, .
	\end{equation}
	The local order parameter of the superfluid phase is related to the pair potential via~\cite{ChamelAllard2020}
	\begin{align}\label{eq:OrderParameter}
		\Psi_q(\rb,t)=\frac{1}{4} \left(\frac{\delta E}{\delta \vert\widetilde{n}_q(\rb,t)\vert^2}\right)^{-1}\D(\rb,t)\, .
	\end{align}
	This order parameter is complex in general and can thus be written as
	\begin{align}
		\Psi_q(\rb,t)=\vert\Psi_q(\rb,t)\vert\exp( i \phi_q(\rb,t))\, .
	\end{align}
	The gradient of its phase $\phi_q(\rb,t)$ defines the superfluid velocity $\pmb{V_q}(\rb,t)$ as follows: 
	\begin{align}\label{eq:SuperfluidVelocities}
		\pmb{V_q}(\rb,t)=\frac{\hbar}{2m_q} \pmb{\nabla}\phi_q(\rb,t)\, .
	\end{align}
	
	The self-consistency of the TDHFB equations is contained in the following  expressions for the particle and pair density matrices: 
	\beqy
	n_q(\rb, \sg; \rp, \sgp;t)=\sum_k &&\left[\f_k \psi_{1k}^{(q)}(\rb,\sg;t ) \psi_{1k}^{(q)}(\rp,\sgp;t)^* \right. \nonumber  \\ 
	&&\left. +(1-\f_k)\sg\sgp\psi_{2k}^{(q)}(\rb,-\sg;t)^* \psi_{2k}^{(q)}(\rp,-\sgp;t ) \right]\, ,
	\eeqy
	\beqy
	\widetilde{n}_q(\rb, \sg; \rp, \sgp ;t)=\sum_k &&\left[\f_k \psi_{1k}^{(q)}(\rb,\sg) \psi_{2k}^{(q)}(\rp,\sgp; t)^* \right. \nonumber \\ 
	&&\left. -(1-\f_k)\sg\sgp\psi_{2k}^{(q)}(\rb,-\sg)^* \psi_{1k}^{(q)}(\rp,-\sgp) \right]\, 
	\eeqy
	with $\f_k$ being the distribution of quasiparticle excitations at temperature $T$.

	\subsection{Application to homogeneous superfluid mixtures with stationary flows}
	\label{sec:TDHFB-neutron-star-cores}
	
	Considering a homogeneous  superfluid mixture with stationary flows in the normal fluid rest frame, the TDHFB equations can be solved exactly~\cite{ChamelAllard2020}. Expressing the pair potential as  $\D(\rb)=\D \exp( 2 i \pmb{Q_q}\cdot \rb)$, where $\D=\vert\D(\rb)\vert$. 
	The order parameter~\eqref{eq:OrderParameter} becomes in this case  
	\begin{align}\label{eq:OrderParameterHom}
		\Psi_q(\rb)=\frac{1}{4} \left(\frac{\delta E}{\delta \vert\widetilde{n}_q\vert^2}\right)^{-1}\D \exp( 2 i \pmb{Q_q}\cdot \rb)\, .
	\end{align}
	The superfluid velocity~\eqref{eq:SuperfluidVelocities} thus reads  $\pmb{V_q}=\hbar\pmb{Q_q}/m_q$. 
	In this stationary situation, the partial derivative $i\hbar \partial/\partial t$ in Eq.~\eqref{eq:TDHFB-Russian} leads to the multiplication by the energy of a quasiparticle excitation with momentum $\hbar \pmb{k}$ given by 
	\begin{align}\label{eq:QuasiparticleEnergy}
		\QuasipartEnergy=\hbar\pmb{k}\cdot\pmb{\Vq} + \sqrt{\varepsilon^{(q)2}_{\pmb{k}} +  \Delta_q^2}\, ,
	\end{align}
	with 
	\begin{align}\label{eq:varepsilon}
		\varepsilon^{(q)}_{\pmb{k}} =\frac{\hbar^2\pmb{k}^2}{2m^{\oplus}_q} +\frac{1}{2}\mqOplus\left(\pmb{\Vq}+\frac{\pmb{I_q}}{\hbar}\right).\left(\pmb{\Vq}-\frac{\pmb{I_q}}{\hbar}\right)+ U_q - \lambda_q \, ,
	\end{align}
	and we have introduced the effective superfluid velocity
	\begin{align}\label{eq:EffectiveSuperfluidVelocity}
		\pmb{\Vq}\equiv \frac{m_q}{\mqOplus}\pmb{V_q}+\frac{\pmb{I_q}}{\hbar}\, .  
	\end{align}
	
	For superfluids at rest (in the normal frame), the lowest possible quasiparticle energy $\QuasipartEnergy$ is finite and given by $\D$, which in this case represents a gap in the quasiparticle energy spectrum. However, in the presence of arbitrary currents, $\D$ does not necessarily imply the existence of a gap. As we shall show, the quasiparticle energy spectrum may be continuous while $\D$ is finite: in this peculiar regime, the nucleons remains superfluid since the order parameter~\eqref{eq:OrderParameterHom} does not vanish. In any case, $\D$ is obtained from the self-consistent equations
	\begin{align}
		\label{eq:GapEquation}
		\D(T,\pmb{\Vq}) = -\frac{2}{V}\frac{\delta E}{\delta\vert \widetilde{n}_q\vert^2} \sum_{\pmb{k}} \frac{\D(T,\pmb{\Vq})}{\sqrt{\varepsilon_{\pmb{k}}^{(q)2} +\D(T,\pmb{\Vq})^2}}\tanh\left(\frac{\beta}{2}\QuasipartEnergy\right)\, ,
	\end{align}
	where $\beta = \left(k_\text{B}T\right)^{-1}$ ($k_\text{B}$ being the Boltzmann constant) and it is understood that the summation must be regularized to remove ultraviolet divergences by means of introducing a cutoff $\Cutoff$ (see, e.g., Ref.~\cite{bulgac2002} for discussions). Equation~\eqref{eq:GapEquation} must be solved together with the particle number conservation
	\begin{align}
		\label{eq:DensityHomogeneous}
		n_q=
		\frac{1}{V}\sum_{\pmb{k}}\left[1-\frac{\varepsilon^{(q)}_{\pmb{k}}}{\sqrt{\varepsilon^{(q)2}_{\pmb{k}} +  \D^2}}\tanh\left(\frac{\beta}{2}\QuasipartEnergy\right)\right]\, . 
	\end{align}
	As can be seen from Eq.~\eqref{eq:varepsilon},  Eqs.~\eqref{eq:GapEquation} and \eqref{eq:DensityHomogeneous} both depend on the \emph{reduced} chemical potential defined by 
	\beqy\label{eq:ReducedChemicalPotential}
	\mu_q = \lambda_q - U_q - \frac{1}{2}\mqOplus\left(\pmb{\Vq}+\frac{\pmb{I_q}}{\hbar}\right).\left(\pmb{\Vq}-\frac{\pmb{I_q}}{\hbar}\right)\, ,
	\eeqy 
	so that $\D$ does not require  the explicit form of the potential $U_q$. 
	
	Each  species can be characterized by the Fermi energy $\Fermi=\hbar^2k_{Fq}^2/(2\mqOplus)$, the Fermi temperature $T_{Fq}=\Fermi/k_\text{B}$ and the Fermi velocity $V_{Fq}=\hbar k_{Fq}/\mqOplus$ (recalling that the Fermi wave-number is given by $k_{Fq}=(3\pi^2 n_q)^{1/3}$). In the following and for convenience, we will introduce the following dimensionless ratios: 
	\beqy\label{eq:BarQuantities}
	\Tbar = \frac{T}{T_{Fq}}\, , \quad \mubar=\frac{\mu_q}{\Fermi}\, ,  \quad \Lbar=\frac{\Cutoff}{\Fermi}\, , \quad \Dbar=\frac{\D}{\Fermi}\, , \quad \Vbar = \frac{\Vq}{V_{Fq}}\, . 
	\eeqy

	We will also take the continuum limit, i.e. we will replace discrete summations over wave vectors by integrations as follows: 
	\beqy\label{eq:continuum}
	\frac{1}{V} \sum_{\pmb{k}} \dotsi \rightarrow   \int \frac{\text{d}^3\pmb{k}}{(2\pi)^3} \dotsi =\int \frac{\text{d}\Omega_{\pmb{k}}}{4 \pi}\int \text{d}\varepsilon\, \mathcal{D}_q(\varepsilon) \dotsi
	\eeqy 
	with $\Omega_{\pmb{k}}$ the solid angle in $\pmb{k}$-space and $\mathcal{D}_q(\varepsilon)$ the density of single-particle states per spin given by 
	\beqy\label{eq:DoS}
	\mathcal{D}_q(\varepsilon)\equiv \int \frac{\text{d}^3\pmb{k}}{(2\pi)^3} \delta(\varepsilon-\varepsilon_{\pmb{k}}^{(q)})=\frac{m_q^\oplus}{2 \pi^2 \hbar^3}\sqrt{2 m_q^\oplus(\varepsilon+\mu_q)}\, ,
	\eeqy 
	and we have made use of Eq.~\eqref{eq:varepsilon}. After integrating over solid angle and changing variables (setting $x=\mubar+\varepsilon/\Fermi=\pmb{k}^2/k_{Fq}^2$), Eq.~\eqref{eq:GapEquation} and the particle number conservation~\eqref{eq:DensityHomogeneous} become, respectively 
	\beqy\label{eq:DimensionlessGapEquation}
	\Delta_q&=& 
	-\frac{1}{4}v^{\pi q} \DqZero \Delta_q \frac{\Tbar}{\Vbar}\displaystyle \int_0^{\mubar + \Lbar}\frac{\text{d}x}{\Ex} \nonumber \\
	&&\qquad\times \log\left[\cosh\left(\frac{\Ex}{2\Tbar} +\frac{\Vbar}{\Tbar}\sqrt{x}\right) \sech\left(\frac{\Ex}{2\Tbar} -\frac{\Vbar}{\Tbar}\sqrt{x}\right)  \right]\, ,
	\eeqy
	\beqy\label{eq:DimensionlessChemicalPotentialEquation}
	\frac{4}{3}&=&\int_0^{+\infty}\text{d}x\;\left\{\sqrt{x}-\frac{\Tbar}{\Vbar}\frac{x-\mubar}{2\Ex} \right. \nonumber \\ 
	&&\qquad\left.\times \log\left[\cosh\left(\frac{\Ex}{2\Tbar} +\frac{\Vbar}{\Tbar}\sqrt{x}\right) \sech\left(\frac{\Ex}{2\Tbar} -\frac{\Vbar}{\Tbar}\sqrt{x}\right)  \right]\right\}\, ,
	\eeqy
	with the following shorthand notation
	\beqy 
	\mathcal{D}_q(0)\equiv\frac{k_{Fq}m^{\oplus}_q}{2\pi^2 \hbar^2}\, , 
	\eeqy 
	and 
	\begin{align}
		\Ex=\sqrt{\left(x-\mubar\right)^2+\Dbar^2}\, ,
	\end{align}
	and we have introduced the pairing strength $v^{\pi q}$ (which may depend on the densities and currents) through the relation 
	\begin{align}
		\frac{\delta E}{\delta\vert \widetilde{n}_q\vert^2} = \frac{1}{4}v^{\pi q}<0\, .
	\end{align}
	
	It is worth remarking that although $\D$ depends in general on the directions of the superfluid velocities $\pmb{V_q}$, this dependence is entirely contained in the absolute value of the 
	effective superfluid velocities $\pmb{\Vq}$. 
	
	\section{Properties of nuclear superfluids in neutron stars}
	\label{sec:properties}
	
	For numerical applications, we will consider $npe\mu$ matter in beta equilibrium, as found in the outer core of neutron stars. The present formalism could also be applied to the neutron superfluid in the inner crust of neutron stars to the extent that the effects of spatial inhomogeneities can be neglected for the temperatures of relevance (see, e.g., Refs.~\cite{chamel2009,chamel2010}). Unless stated otherwise, we will adopt the Brussels-Montreal functional BSk24~\cite{goriely2013} for which the equation of state throughout all regions of a neutron star is available~\cite{pearson2018,pearson2020,pearson2022}.  This equation of state is consistent with astrophysical constraints coming from analyses of the gravitational-wave signal from the binary neutron-star merger GW170817 and of its electromagnetic counterpart~\cite{perot2019}. Moreover, we have already employed this functional in our previous study to calculate some superfluid properties in neutron-star cores~\cite{ChamelAllard2021}. The functional BSk24 was constructed 
	by fitting directly effective masses and $^1$S$_0$ pairing gaps, as obtained from diagrammatic many-body 
	calculations based on realistic two- and three-body interactions~\cite{cao2006}. 
	
	\subsection{Critical temperature}
	
	In the absence of currents, Eq.~\eqref{eq:DimensionlessGapEquation} for $\D$, which is interpretable as the pairing gap under such conditions,  
	becomes  
	\begin{align}\label{eq:DimensionlessGapEquation0}
		1= -\frac{1}{2} v^{\pi q}\DqZero\displaystyle \int_0^{\mubar + \Lbar}\text{d}x\frac{\sqrt{x}}{\Ex}  \tanh\left(\frac{\Ex}{2\Tbar}\right)\, . 
	\end{align}
	At $T=0$, Eq.~\eqref{eq:DimensionlessGapEquation0} further reduces to 
	\beqy\label{eq:DimensionlessGapEquation00}
	1=-\frac{1}{2}  v^{\pi\, q}\DqZero \mathcal{I}_q\biggl[\mu_q(T=0,\Vq=0);\Delta_q(T=0,\Vq=0)\biggr]\, ,
	\eeqy 
	where we have introduced the integral 
	\beqy\label{eq:Integral1}
	\mathcal{I}_q\biggl[\mu_q(T,\Vq);\Delta_q(T,\Vq)\biggr] \equiv \displaystyle \int_0^{\mubar + \Lbar}\text{d}x\frac{\sqrt{x}}{\Ex}  = \int_0^{\mubar+\Lbar} \text{d}x \sqrt{\frac{x}{
			\left(x-\mubar\right)^2+\Dbar^2}}\, .
	\eeqy
	
	In the weak-coupling approximation, i.e. $\Delta_q\ll \Fermi,\Cutoff$, it is a very good approximation to neglect $\Delta_q$ in  Eq.~\eqref{eq:DimensionlessChemicalPotentialEquation} so that $\mu_q\approx \Fermi $. As shown in Ref.~\cite{Chamel2010Pairing}, the integral~\eqref{eq:Integral1} is approximately given by 
	\beqy \label{eq:Integral1App}
	\mathcal{I}_q\approx 2\log\left(\frac{2}{\Dbar}\right)+ \log (16 \Lbar) + 2\sqrt{1+\Lbar}-2\log\left(1+\sqrt{1+\Lbar}\right)-4 \, .
	\eeqy 
	Solving for $\D$ yields 
	\beqy\label{eq:ChamelApprox}
	\Delta^{(0)}_q \equiv \Delta_q(T=0,\Vq=0) \approx  \frac{8\sqrt{\Fermi\Cutoff}}{1+\sqrt{1+\Cutoff/\Fermi}}\exp\left[\frac{1}{v^{\pi q}\mathcal{D}_q(0)} + \sqrt{1+\frac{\Cutoff}{\Fermi}}-2\right]\, .
	\eeqy 
	It should be stressed that this expression was obtained by going beyond the usual (zeroth-order) ``weak-coupling approximation'' according to which the density of single-particle states  $\mathcal{D}_q(\varepsilon)$ is taken outside the integral and evaluated at the Fermi level. Even though 
	this provides a good approximation in the case of conventional Bardeen-Cooper-Schrieffer (BCS)  superconductivity~\cite{Bardeen1957}, it is less accurate in the nuclear context, especially at low densities, 
	because essentially all states lying below the cutoff are involved in the pairing mechanism. 
	
	We will now follow the same line of reasoning as in Ref.~\cite{Chamel2010Pairing} to estimate the critical temperature $\Tcq$ at which the pairing gap vanishes, i.e. $\Delta_q(T=T^{(0)}_{cq},\Vq=0)=0$. 
	In this case, the gap equation~\eqref{eq:DimensionlessGapEquation0} assuming  $\mu_q\approx \Fermi$ reduces to
	\beqy\label{eq:GapEquationTcq}
	1=-\frac{1}{2} v^{\pi\, q}\DqZero \,  \mathcal{J}_q \, ,
	\eeqy
	with
	\beqy\label{eq:Integral2}
	\mathcal{J}_q\equiv \int_{-1/2\bar{T}^{(0)}_{cq}}^{\Lbar /2\bar{T}^{(0)}_{cq}} {\rm d} u  \frac{\tanh|u|}{|u|}\sqrt{1+2u \bar{T}^{(0)}_{cq}}\, .
	\eeqy
	Expanding the square root in the integrand in powers of $\bar{T}^{(0)}_{cq}$ leads to
	\beqy
	\mathcal{J}_q=\sum_{r=0}^{+\infty}\frac{(2r)!}{(1-2r)(r!)^2}\left(-\frac{\bar{T}_{cq}^{(0)}}{2}\right)^r
	\int_{-1/2\bar{T}_{cq}}^{\Lbar/2\bar{T}_{cq}}{\rm d}u\, \frac{u^r}{|u|}\tanh|u| \, .
	\eeqy
	The integral $\mathcal{J}_q$ is generally evaluated by keeping only the first term. 
	Using the approximation
	\beqy
	\int_0^y {\rm d}u\frac{\tanh u}{u} \approx \log\left(\frac{4y}{\pi}\right) + \gamma \, ,
	\eeqy
	for $y\gg 1$ ($\gamma\simeq 0.57722$ being the Euler-Mascheroni constant), we find
	\beqy\label{eq:Jqtilde0}
	\mathcal{J}_q \approx 
	\int_{-1/2\bar{T}^{(0)}_{cq}}^{\Lbar/2\bar{T}^{(0)}_{cq}}{\rm d}u\, \frac{\tanh|u|}{|u|}\approx \log\left[\frac{4 \Lbar}{(\bar{T}^{(0)}_{cq})^{2} \pi^2}\right] +2\gamma\, .
	\eeqy
	
	It is not so much difficult to evaluate the higher-order coefficients. 
	Calculating the integrals, 
	keeping as before the leading terms, and summing all coefficients yield 
	\beqy\label{eq:sum-Jqtilde-exact}
	\mathcal{J}_q=&&2\gamma +\log\left[\frac{4\Fermi\Cutoff}{(k_\text{B}T^{(0)}_{cq})^{2} \pi^2}\right] \biggr. \nonumber \\ 
	&&+2\sqrt{1+\frac{\Cutoff}{\Fermi}}-4 +\log 16 - 2 \log\left(1+\sqrt{1+\frac{\Cutoff}{\Fermi}}\right)\, .
	\eeqy
	Substituting Eq.~\eqref{eq:sum-Jqtilde-exact} in Eq.~\eqref{eq:GapEquationTcq} and solving for $T^{(0)}_{cq}$ 
	leads to the familiar scaling relation
	\beqy\label{eq:Tc0}
	\frac{k_\text{B} T^{(0)}_{cq}}{\Delta_q^{(0)}}= \frac{\exp(\gamma)}{\pi}\simeq 0.56693 \, .
	\eeqy
	Note that this relation was historically obtained in the crudest version of the weak-coupling approximation
	according to which the density of single-particle states is taken as a constant in the integral, i.e. only the first term $r=0$ is retained while all higher-order terms 
	for $r\geq1$ are  dropped~\cite{Bardeen1957}. We have thus shown that the validity of Eq.~\eqref{eq:Tc0} is more general and extends well beyond the first term $r=0$, 
	in fact it holds to \emph{all} terms. 
	
	\subsection{Critical velocities}
	\label{subsec:critical-velocities}
	
	Substituting Eq.~\eqref{eq:DimensionlessGapEquation00} into \eqref{eq:DimensionlessGapEquation}, the equation for $\D$ can be alternatively written as
	\begin{align}\label{eq:AlternativeDimensionlessGapEquation}
		&\mathcal{I}_q\biggl[\mu_q(T=0,\Vq=0);\Delta_q(T=0,\Vq=0)\biggr]-\mathcal{I}_q\biggl[\mu_q(T,\Vq);\Delta_q(T,\Vq)\biggr] \nonumber \\ 
		&=\int_0^{\mubar + \Lbar}\frac{\text{d}x}{\Ex}  \biggl\{ \log\left[\cosh\left(\frac{\Ex}{2\Tbar} +\frac{\Vbar}{\Tbar}\sqrt{x}\right)  \sech\left(\frac{\Ex}{2\Tbar} -\frac{\Vbar}{\Tbar}\sqrt{x}\right)  \right]\frac{\Tbar}{2\Vbar} -\sqrt{x}\biggr\} \, .
	\end{align} 
	If the superfluid velocity is small enough such that $\Ex > 2 \Vbar \sqrt{x}$  for all $x$ between 0 and $\mubar+\Lbar$, the  argument of the logarithm in the  right-hand side of Eq.~\eqref{eq:AlternativeDimensionlessGapEquation} is approximately given by 
	\beqy 
	\cosh\left(\frac{\Ex}{2\Tbar} +\frac{\Vbar}{\Tbar}\sqrt{x}\right)  \sech\left(\frac{\Ex}{2\Tbar} -\frac{\Vbar}{\Tbar}\sqrt{x}\right) \approx  \exp\left( \frac{2\Vbar}{\Tbar}\sqrt{x}\right)
	\eeqy 
	at low temperatures $\Tbar \ll 1$. 
	Taking the limit $T=0$ of Eq.~\eqref{eq:AlternativeDimensionlessGapEquation} thus leads to 
	\beqy 
	\mathcal{I}_q\biggl[\mu_q(T=0,\Vq=0);\Delta_q(T=0,\Vq=0)\biggr]=\mathcal{I}_q\biggl[\mu_q(T=0,\Vq);\Delta_q(T=0,\Vq)\biggr] \, . 
	\eeqy 
	Similarly, at $T=0$ Eq.~\eqref{eq:DimensionlessChemicalPotentialEquation} reduces to
	\beqy
	\frac{4}{3}&=&\int_0^{+\infty}\text{d}x\;\sqrt{x}\left(1-\frac{x-\mubar}{\Ex} \right)\, .
	\eeqy
	This shows that $\mu_q(T=0,\Vq)$ depends implicitly on $\Vq$ through $\Delta_q(T=0,\Vq)$ contained in $\Ex$. As a consequence, both $\mu_q(T=0,\Vq)=\mu_q(T=0,\Vq=0)$ and $\Delta_q(T=0,\Vq)=\Delta_q(T=0,\Vq=0)$ are independent of the superflows. This conclusion however does not hold for arbitrarily large superfluid velocity $\Vq$. Indeed,  the integral  on the right-hand side of  Eq.~\eqref{eq:AlternativeDimensionlessGapEquation} no longer vanishes if $\Ex \leq 2 \Vbar \sqrt{x}$ for some $x$. 
	
	Let us examine the behavior of $\Ex - 2 \Vbar \sqrt{x}$. For this quantity to be negative, we must have 
	\beqy 
	(x-\mubar)^2-4 \Vbar^2 x +\Dbar^2 \leq 0 \, .
	\eeqy 
	It can be easily shown  that this condition is fulfilled if  $x$ lies  between $x_-$ and $x_+$ defined by
	\beqy \label{eq:xpm}
	x_\pm = \mubar +2\Vbar^2 \pm 2 \sqrt{\mubar \Vbar^2 +\Vbar^4-\frac{1}{4}\Dbar^2} \, ,
	\eeqy 
	provided $4\mubar \Vbar^2 +4\Vbar^4-\Dbar^2\geq 0$. 
	For this inequality to hold, the effective  superfluid velocity must exceed some critical value $\VLq $ given by 
	\beqy\label{eq:Landau-effective-velocity}
	\bar{\mathbb{V}}_{Lq} \equiv \sqrt{\frac{\mubar}{2}\Biggl[ \sqrt{1+\left(\frac{\Dbar}{\mubar}\right)^2}-1\Biggr] }
	\eeqy 
	recalling that both $\mubar$ and $\Dbar$ are independent of $\Vq$ for $\Vq \leq \VLq$. Let us remark that $\Ex - 2 \Vbar \sqrt{x}$ is nothing but the lowest value of the quasiparticle energy~\eqref{eq:QuasiparticleEnergy} in units of the Fermi energy, 
	obtained for the wavevector $\pmb{k}$ anti-parallel to $\pmb{\Vq}$.
	When $\mathbb{V}_{q}<\VLq$, $\Ex - 2 \Vbar \sqrt{x}$ hence also $\QuasipartEnergy$ remain always strictly positive, i.e. there exists a 
	gap in the quasiparticle energy spectrum. This is no longer the case for $\mathbb{V}_{q}=\VLq$. In other words, this velocity corresponds to 
	the vanishing of the quasiparticle gap even though $\D$ remains finite. 
	
	The velocity $\VLq$ coincides with Landau's critical velocity, here derived for the case of strongly interacting nuclear superfluid mixtures. Indeed, 
	Eq.~\eqref{eq:Landau-effective-velocity} can be alternatively obtained using Landau's criterion according to which the critical velocity is determined by the minimum of $\mathfrak{E}_{k}^{(q)}/(\hbar k)$, where the quasiparticle energy $\mathfrak{E}_{k}^{(q)}$ is here evaluated for the superfluids at rest in the normal frame. However, it should be stressed that Landau's criterion applies here to the \emph{effective} rather than the true superfluid velocities $\pmb{V_q}$. 
	Due to mutual entrainment effects between neutrons and protons, $\VLq$ will generally depend on both $\pmb{V_n}$ and $\pmb{V_p}$. 
	The expression~\eqref{eq:Landau-effective-velocity} also generalizes that obtained earlier in the context of cold atoms for a single superfluid within the BCS theory using a much simpler Hamiltonian including only a kinetic term~\cite{combescot2006}. In the weak-coupling approximation $\mubar \approx 1$ and $\Dbar \ll 1$, $\VLq$ reduces to lowest order to 
	\beqy\label{eq:Landau-velocity-approx}
	\VLq \approx \frac{\D(T=0,\Vq=0)}{\hbar k_{Fq}} = \frac{\Delta^{(0)}_q}{\hbar k_{Fq}}  \, .
	\eeqy 
	In the limit of a single constituent, this expression coincides with the well-known expression from the BCS theory of electron superconductivity~\cite{Parmenter1962,bardeen1962} recalling that $\mathbb{V}_{q}=V_q$ in this case~\cite{ChamelAllard2021}. Our analysis shows that the same expression still holds for mixtures but in terms of the \emph{effective} superfluid velocities.

	Let us assume that  $\mathbb{V}_{q} \geq \VLq $. At low temperatures $\Tbar \ll 1$,  the  argument of the logarithm in the  right-hand side of Eq.~\eqref{eq:AlternativeDimensionlessGapEquation} is now approximately given by
	\beqy 
	\cosh\left(\frac{\Ex}{2\Tbar} +\frac{\Vbar}{\Tbar}\sqrt{x}\right)  \sech\left(\frac{\Ex}{2\Tbar} -\frac{\Vbar}{\Tbar}\sqrt{x}\right) \approx  \exp\left( \frac{\Ex}{\Tbar}\right) \,  ,
	\eeqy 
	if $\Ex \leq 2 \Vbar \sqrt{x}$ (i.e. for $x$ values between $x_-$ and $x_+$). 
	Taking the limit $T=0$, Eq.~\eqref{eq:AlternativeDimensionlessGapEquation} becomes\footnote{Let us recall that the contribution to the right-hand side of Eq~\eqref{eq:AlternativeDimensionlessGapEquation} from $x$ values outside the $[x_{-};x_{+}]$ interval has been shown to vanish identically. }
	\begin{align}
		&\mathcal{I}_q\biggl[\mu_q(T=0,\Vq=0);\Delta_q(T=0,\Vq=0)\biggr]-\mathcal{I}_q\biggl[\mu_q(T,\Vq);\Delta_q(T,\Vq)\biggr] \nonumber \\ 
		&=\int_{x_-}^{x_+}\frac{\text{d}x}{\Ex} \left(\frac{\Ex}{2\Vbar}  -\sqrt{x}\right) \nonumber \\ 
		&=\frac{x_+-x_-}{2 \Vbar} - \int_{x_-}^{x_+}\text{d}x \sqrt{\frac{x}{(x-\mubar)^2+\Dbar^2} } \, .
	\end{align} 
	Using the  definition~\eqref{eq:Integral1}, 
	this equation  can be equivalently written as 
	\begin{align}\label{eq:AlternativeDimensionlessGapEquationBis} 
		&\mathcal{I}_q\biggl[\mu_q(T=0,\Vq=0);\Delta_q(T=0,\Vq=0)\biggr] \nonumber \\ 
		&=\frac{x_+-x_-}{2 \Vbar} + \int_{0}^{x_-}\text{d}x \sqrt{\frac{x}{(x-\mubar)^2+\Dbar^2} } + \int_{x_+}^{\mubar+\Lbar}\text{d}x \sqrt{\frac{x}{(x-\mubar)^2+\Dbar^2} }  \, .
	\end{align}
	This equation admits nontrivial solutions $\D\neq 0$ even though $\Vq>\VLq$. However, $\D$  is no longer independent of the superflow but decreases with increasing $\Vq$.  Superfluidity thus disappears at some critical velocity $\mathbb{V}^{(0)}_{cq}>\VLq$ such that $\Psi_q(T=0,\Vq=\Vcq)=0$ therefore $\D(T=0,\Vq=\Vcq)=0$ in view of Eq.~\eqref{eq:OrderParameterHom}. To estimate this velocity let us set $\Dbar=0$ and $\mubar=1$. The roots  $x_\pm$  reduce to $ 1 + 2 \Vbar^2 \pm 2 \Vbar \sqrt{1 + \Vbar^2}$. Assuming $x_+< 1+\Lbar$ and integrating, we find 
	\begin{align}
		&\mathcal{I}_q\biggl[\mu_q(T=0,\Vq=0);\Delta_q(T=0,\Vq=0)\biggr] \nonumber \\ 
		&=2 \sqrt{1 + \Vbar^2} + \int_{0}^{x_-}\text{d}x \frac{\sqrt{x}}{1-x}  + \int_{x_+}^{1+\Lbar}\text{d}x \frac{\sqrt{x}}{x-1}   \nonumber \\ 
		&=2 \sqrt{1 + \Vbar^2}   -2 \sqrt{x_-} + 2 \, \mathrm{arctanh}\sqrt{x_-} 
		+2 (\sqrt{1+\Lbar} - \sqrt{x_+}) \nonumber \\ 
		&\qquad +2  \left( \mathrm{arcoth} \sqrt{x_+}-\mathrm{arcoth} \sqrt{1+\Lbar} \right)
		\, .
	\end{align}
	Expanding the right-hand side to lowest order in $\Vbar\ll1$, we obtain 
	\begin{align}
		&\mathcal{I}_q\biggl[\mu_q(T=0,\Vq=0);\Delta_q(T=0,\Vq=0)\biggr] \nonumber \\ 
		&\approx 
		2  \Bigl(    \sqrt{1 +\Lbar} -1 - \mathrm{arcoth} \sqrt{1+\Lbar} + \log 2 -  \log \Vbar \Bigr)
		\, .
	\end{align}
	Substituting Eq.~\eqref{eq:Integral1App} and solving  for $\Vbar$ yields 
	\beqy 
	\bar{\mathbb{V}}^{(0)}_{cq}\approx \frac{\Euler}{4}  \Dbar(T=0,\Vq=0) \, ,
	\eeqy 
	or  equivalently 
	\beqy \label{eq:Vcq}
	\mathbb{V}^{(0)}_{cq}\approx \frac{\Euler}{2} \frac{\D(T=0,\Vq=0)}{\hbar k_{Fq}} = \frac{\Euler}{2} \VLq \approx 1.35914 \VLq \, ,
	\eeqy 
	where $\Euler \approx 2.71828$ is Euler's number and we have made use of Eq.~\eqref{eq:Landau-velocity-approx}. 
	In the limit of a single constituent, this expression coincides with Eq.~(30) of Ref.~\cite{Parmenter1962} 
	obtained in the context of conventional BCS electron superconductivity. However, the present derivation shows 
	that this result is quite general and remains valid (i) for a more general class of single-particle Hamiltonians $h_q(\pmb{r})$ including not only a kinetic term but involving also a scalar potential $U_q$ as well as a vector potential $\pmb{I_q}$, (ii) for mixtures provided $\mathbb{V}^{(0)}_{cq}$ are understood as \emph{effective} superfluid velocities. 
	
	In the absence of superflows, the quasiparticle energy spectrum at zero temperature exhibits a gap given by $\Delta_q^{(0)}$, as can be seen from Eq.~\eqref{eq:QuasiparticleEnergy}. With increasing effective superfluid velocity $\Vq$, the gap shrinks as previously discussed and vanishes for $\Vq = \VLq$, 
	recalling that $\Delta_q(T=0,\Vq\leq \VLq)=\Delta_q^{(0)}$ remains unchanged. At still higher effective superfluid velocities $\Vq >  \VLq$ and  $\Vq < \mathbb{V}^{(0)}_{cq}$, the quasiparticle energies $\QuasipartEnergy$ thus form a continuum without any gap even though the order parameter therefore $\Delta_q(T=0,\Vq)$ itself does not vanish: the superfluid phase is \emph{gapless}. Landau's criterion implies the existence of quasiparticle excitations for $\Vq >  \VLq$ even at $T=0$. Indeed, the quasiparticle distribution at any temperature $T$ is given by~\cite{blaizot1986}
	\begin{equation}\label{eq:QuasiparticleDistribution}
		f^{(q)}_{\pmb{k}} = \left[1+\exp\left(\beta \QuasipartEnergy\right)\right]^{-1}=\frac{1}{2}\left[1-\tanh\left(\frac{\beta}{2} \QuasipartEnergy\right)\right]\, .
	\end{equation}
	At $T=0$, we thus have $f^{(q)}_{\pmb{k}}=0$ for any $\pmb{k}$ if $\Vq <  \VLq$ (since $\QuasipartEnergy>0$). If $\Vq >  \VLq$, $f^{(q)}_{\pmb{k}}=1$ for some $\pmb{k}$ for which  $\QuasipartEnergy<0$.  Recalling that the entropy density within the HFB approach is given by (see, e.g. Ref.~\cite{blaizot1986})
	\beqy \label{eq:Entropy}
	s_q=-\frac{2 k_{\rm B}}{V} \sum_{\pmb{k}} \left[f^{(q)}_{\pmb{k}}\log f^{(q)}_{\pmb{k}} +(1-f^{(q)}_{\pmb{k}})\log(1-f^{(q)}_{\pmb{k}})\right]\, ,
	\eeqy 
	where $k_{\rm B}$ denotes Boltzmann's constant, the third law of thermodynamics requiring $s_q$ to vanish thus remains fulfilled despite the presence of 
	quasiparticle excitations. The gapless phase only exists for effective superfluid velocities $\Vq \geq  \VLq$ and  $\Vq < \mathbb{V}^{(0)}_{cq}$. 
	
	The critical effective superfluid velocity above which superfluidity is destroyed at any given temperature is well fitted by the following expression~\cite{ChamelAllard2021}: 
	\beqy
	\displaystyle
	\mathbb{V}_{cq}(T\leq \Tcq)\simeq \Vcq \sqrt{1-\left(\frac{T}{\Tcq}\right)^{2.508}}\, . 
	\eeqy
	The error decreases with increasing density, the maximum relative error is attained at the crust-core boundary but does not exceed 0.95$\%$. 
	The modulus of the order parameter can be represented by the following analytical expression~\cite{ChamelAllard2021}: 
	\beqy\label{eq:GapInterpolation}
	\frac{\vert\Psi_q(T=0,\VLq<\Vq \leq \mathbb{V}^{(0)}_{cq})\vert}{\vert\Psi_q(T=0,\Vq=0)\vert}=
	\frac{\Delta_q(T=0,\VLq<\Vq \leq \mathbb{V}^{(0)}_{cq})}{\Delta_q^{(0)}} \nonumber \\
	=0.5081\sqrt{1-\frac{\Vq}{\mathbb{V}^{(0)}_{cq}}}\left(3.312\frac{\Vq}{\mathbb{V}^{(0)}_{cq}}-3.811\sqrt{\frac{ \mathbb{V}^{(0)}_{cq}}{\Vq}} +5.842\right)\, .
	\eeqy
	The maximum relative error amounts to 0.13\%.

	\subsection{Density of quasiparticle  states}
	\label{sec:density-quasiparticle-states}
	
	The different superfluid regimes can be directly seen by calculating the density of quasiparticle states per  spin defined by
	\beqy\label{eq:DoS-Def}
	\DoS=
	\int \frac{\text{d}^3\pmb{k}}{(2\pi)^3}\delta(\mathcal{E}-\QuasipartEnergy)\, .
	\eeqy
	At low temperatures $\Tbar\ll 1$ and since $\bar{\Vq} \ll 1$, we can make use of Landau's approximations (see Section 2.5 of Ref.~\cite{ChamelAllard2021} for more details). Substituting Eq.~\eqref{eq:QuasiparticleEnergy} in Eq.~\eqref{eq:DoS-Def}, the density of quasiparticle states can thus be approximated by the following integral
	\beqy\label{eq:DoS-Approx}
	\DoS\approx \DqZero\int_{-1}^{1}\text{d} \eta \frac{\mathcal{E}-\hbar k_{Fq}\Vq\eta}{\sqrt{(\mathcal{E}-\hbar k_{Fq}\Vq\eta)^2-\Delta_q^2}}H(\mathcal{E} - \hbar k_{Fq}\Vq\eta - \Delta_q)\, ,
	\eeqy
	where $H$ denotes the Heaviside distribution. 
	Carrying out the integration yields
	\begin{align}\label{eq:QuasiparticleDoS}
		\DoS &= \frac{\DNq}{2\hbar k_{Fq}\Vq} H(\mathcal{E}+\hbar k_{Fq}\Vq-\Delta_q)\left[\sqrt{(\mathcal{E}+\hbar k_{Fq}\Vq)^2-\Delta_q^2}\right.\notag\\
		&\qquad\qquad -H(\mathcal{E}-\hbar k_{Fq}\Vq-\Delta_q) \left.\sqrt{(\mathcal{E}-\hbar k_{Fq}\Vq)^2-\Delta_q^2}\right]\, ,
	\end{align}
	where we have introduced the density of quasiparticle states in the normal phase
	\beqy 
	\DNq\equiv 2\DqZero=\frac{k_{Fq}\mqOplus}{\pi^2 \hbar^2}\, .
	\eeqy 
	
\begin{figure}
		\includegraphics[width=10.5cm]{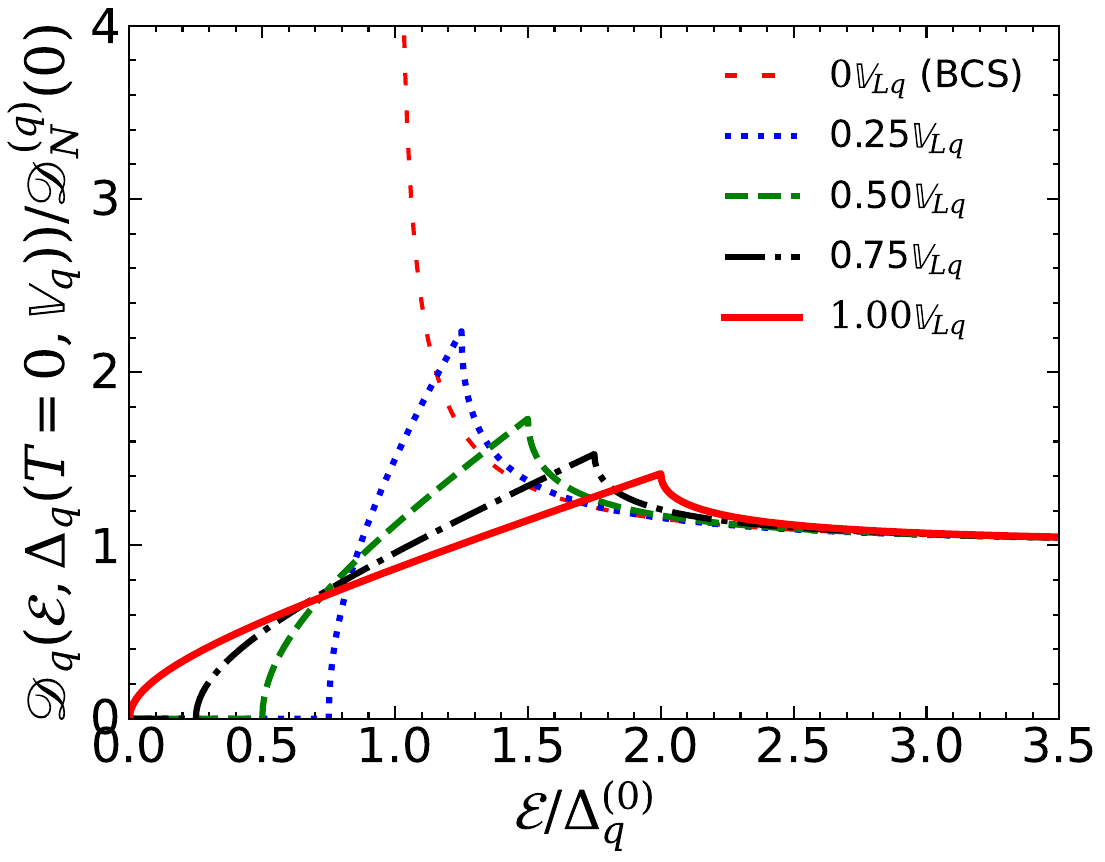}
		\caption{Density of quasiparticle states $\mathscr{D}_q(\mathcal{E},\Delta_q(T=0,\Vq))$ at zero temperature (normalized by $\DNq$) for different effective superfluid velocities up to Landau's velocity $\VLq$. The red line corresponds to the BCS case, for which no quasiparticle state exists for energies $\mathcal{E}\leq\Delta_q^{(0)}$. The presence of a non-vanishing effective superfluid velocity $\Vq$ reduces the value of the gap which disappears at $\VLq$. See text for details.}
		\label{fig:DoS-0toVLq}
	\end{figure}

\begin{figure}
		\includegraphics[width=10.5cm]{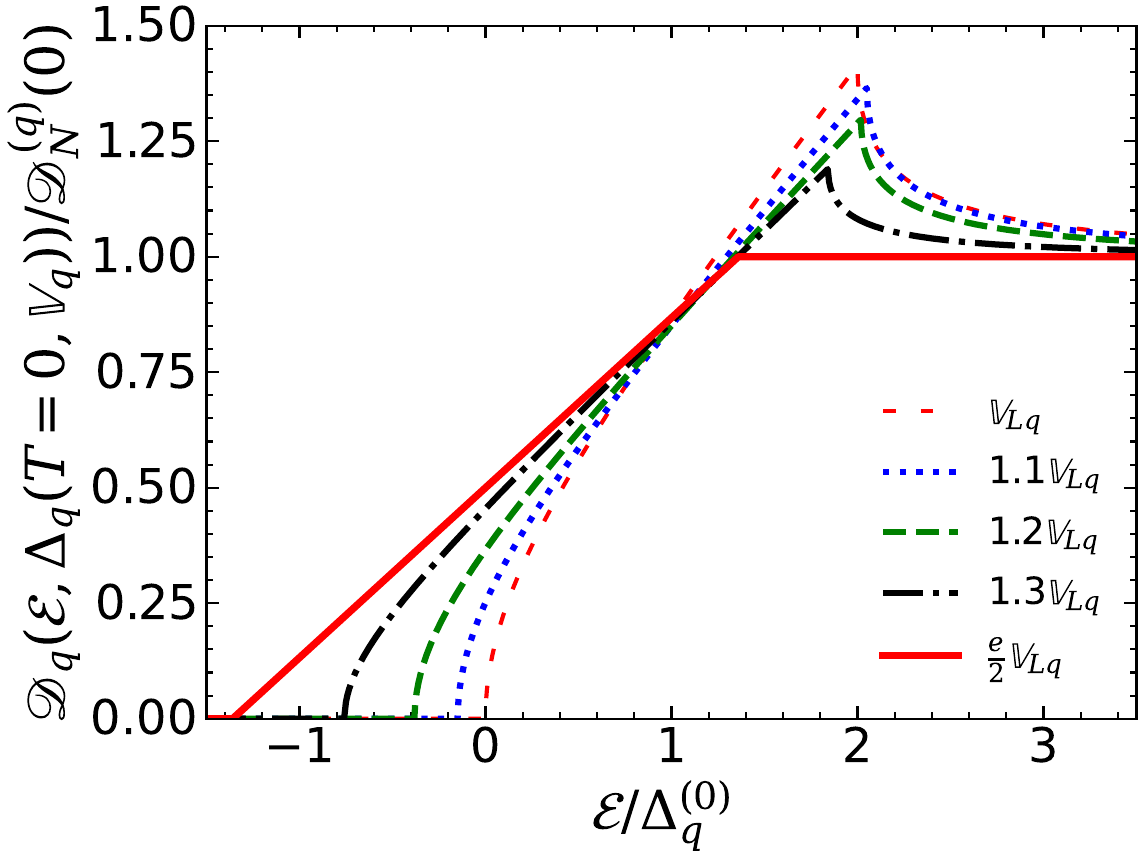}
		\caption{Same as Figure~\ref{fig:DoS-0toVLq} for effective superfluid velocities ranging from Landau's velocity $\VLq$ to the critical velocity $\Vcq$. For $\Vq>\VLq$, quasiparticle states with negative energies $\mathcal{E}$ appear. The superfluid is gapless although the order parameter $\Delta_q$ remains finite.}
		\label{fig:DoS-VLqtoVcq}
	\end{figure}

	In the absence of superflows, the previous expression reduces to
	\beqy\label{eq:DoS-BCS}
	\mathscr{D}_q^{\text{(BCS)}}\left(\mathcal{E},\Delta_q(T,\Vq=0)\right)= \frac{\DNq\mathcal{E}}{\sqrt{\mathcal{E}^2-\Delta_q(T,\Vq=0)^2}}H(\mathcal{E}-\Delta_q(T,\Vq=0))\, . 
	\eeqy
	This expression generalizes the well-known BCS result for a conventional superconductor to a superfluid mixture. 
	The density of quasiparticle states is then identically zero  for quasiparticles energies $\mathcal{E}\leq \Delta_q(T,\Vq=0)$ thus reflecting the existence of a gap in the quasiparticle energy spectrum whenever $\Delta_q\neq 0$ and $\Vq=0$. However, as shown in Fig.~\ref{fig:DoS-0toVLq}, the gap is progressively filled with quasiparticle states as the effective superfluid velocity $\Vq$ is increased, and disappears when $\Vq=\VLq$ while $\Delta_q$ remains unchanged. With further increase of $\Vq$, $\Delta_q$ decreases and eventually vanishes at the critical velocity $\Vq=\Vcq$ 
	while quasiparticle states with negative energies appear, as can be seen in Fig.~\ref{fig:DoS-VLqtoVcq}. 
	
	\subsection{Specific heat}
	
	The disappearance of the gap in the quasiparticle energy spectrum is expected to have a very strong impact on thermal properties, such as the specific heat. 
	Using the expression~\eqref{eq:Entropy} for the entropy density and \eqref{eq:QuasiparticleDistribution} for the quasiparticle distribution functions, 
	the specific heat can be expressed as 
	\beqy\label{eq:SpecificHeatDef}
	c_V^{(q)} = T\frac{\partial s_q}{\partial T} = \frac{k_{\text{B}}}{2}\beta^2 
	\int \frac{\text{d}^3\pmb{k}}{(2\pi)^3} \left(\mathfrak{E}_{\pmb{k}}^{(q) 2} +\beta\QuasipartEnergy\frac{\partial\QuasipartEnergy}{\partial \beta}\right)\sech^2\left(\frac{\beta}{2}\QuasipartEnergy\right) \, .
	\eeqy
	At low temperatures $\beta\Delta_q^{(0)}\gg 1$, the term involving the partial derivative $\partial\QuasipartEnergy/\partial\beta $
	can be neglected. Introducing the density of quasiparticles states and changing variables, the specific heat can thus be written as 
	\beqy\label{eq:SpecificHeat-LowT}
	c_V^{(q)} 
	\approx \frac{k_{\text{B}}}{2\beta}\int_{-\infty}^{+\infty} \text{d}x\; \mathscr{D}_q\left(\frac{x}{\beta},\Delta_q\right)x^2\sech^2\left(\frac{x}{2}\right)\, .
	\eeqy
	
	\subsubsection{Normal phase}
	
	In the normal phase, nucleons obviously remain at rest in the normal fluid frame therefore we must have    $\Vq=0$. Setting $\Delta_q=0$ in Eq.~\eqref{eq:DoS-BCS} and substituting in Eq.~\eqref{eq:SpecificHeat-LowT} leads to 
	the classical result (see, e.g., Ref.~\cite{Abrikosov} in the context of metals)
	
	\beqy\label{eq:CV-normal}
	c_{N}^{(q)}(T^{(0)}_{cq}< T\ll T_{Fq})\approx\frac{\pi^2}{3}\DNq  k_{\text{B}}^2 T=\frac{1}{3}\frac{k_{Fq}\mqOplus}{\hbar^2} k_{\text{B}}^2 T \, . 
	\eeqy
	
	\subsubsection{Superfluid phase: general case}
	
	Substituting Eq.~\eqref{eq:QuasiparticleDoS} in Eq.~\eqref{eq:SpecificHeat-LowT} yields the general expression of the specific heat in the superfluid phase for arbitrary effective superfluid velocities: 
	\begin{align}\label{eq:SpecificHeat-LowT-Explicit}
		c_V^{(q)}(T\ll T^{(0)}_{cq},\Vq) &\approx \frac{k_{\text{B}}\DNq}{4\hbar k_{Fq}\beta\Vq}\int_{\beta(\Delta_q-\hbar k_{Fq}\Vq)}^{+\infty} \text{d}x\;\sqrt{\left(\frac{x}{\beta}+\hbar k_{Fq}\Vq\right)^2-\Delta_q^2}\; x^2\sech^2\left(\frac{x}{2}\right)\notag \\
		&\qquad -\frac{k_{\text{B}}\DNq}{4\hbar k_{Fq}\beta\Vq}\int_{\beta(\Delta_q+\hbar k_{Fq}\Vq)}^{+\infty} \text{d}x\;\sqrt{\left(\frac{x}{\beta}-\hbar k_{Fq}\Vq\right)^2-\Delta_q^2}\; x^2\sech^2\left(\frac{x}{2}\right)\notag\\
		&\approx \frac{3}{4\pi^2}\Cnorm\frac{\VLq}{\Vq}\int_{-\NegativeBound}^{+\infty}\text{d}x \sqrt{\left(\frac{\expgamma}{\pi}\frac{T}{\Tcq}x+\frac{\Vq}{\VLq}\right)^2-\left(\frac{\Delta_q}{\Delta_q^{(0)}}\right)^2}x^2\sech^2\left(\frac{x}{2}\right)\notag\\
		&\qquad - \frac{3}{4\pi^2}\Cnorm\frac{\VLq}{\Vq}\int_{\PositiveBound}^{+\infty}\text{d}x \sqrt{\left(\frac{\expgamma}{\pi}\frac{T}{\Tcq}x-\frac{\Vq}{\VLq}\right)^2-\left(\frac{\Delta_q}{\Delta_q^{(0)}}\right)^2}x^2\sech^2\left(\frac{x}{2}\right)\, ,
	\end{align}
	where the lower bounds of the integrals are defined by 
	\begin{align}\label{eq:Bounds}
		\delta_q^{\pm} = \beta\left(\hbar k_{Fq}\Vq \pm \Delta_q\right) = \frac{\pi}{\expgamma}\frac{\Tcq}{T}\left( \frac{\Vq}{\VLq}\pm \frac{\Delta_q}{\Delta_q^{(0)}}\right)\, .
	\end{align}
	This shows that the ratio $c_{V}^{(q)}/\Cnorm$ is a universal function of $T/\Tcq$, $\Vq/\VLq$, and $\Delta_q/\Delta_q^{(0)}$. 
	Explicit analytical expressions will be derived in the next subsections.

	\subsubsection{Superfluid phase: BCS limit $\Vq=0$}
	
	In the superfluid phase and in the absence of superflows, the specific heat can be easily calculated from 
	Eq.~\eqref{eq:SpecificHeat-LowT} using Eq.~\eqref{eq:DoS-BCS} leading to:
	\beqy\label{eq:BCS-SpecificHeat}
	c_V^{(q)}(T\ll T^{(0)}_{cq},\Vq=0)\approx \frac{3\sqrt{2}}{\pi^{3/2}} \left(\frac{T^{(0)}_{cq}}{T}\frac{\pi}{\expgamma}\right)^{5/2}\exp\left(-\frac{\Tcq}{T}\frac{\pi}{\expgamma}\right)\ \Cnorm(T) \, .
	\eeqy
	The exponential suppression of the specific heat at low temperatures is a well-known prediction of the BCS theory (see, e.g., Ref.~\cite{Abrikosov}) and is a direct consequence of the existence of a gap in the quasiparticle energy spectrum, energies $\mathfrak{E}\leq \Delta_q^{(0)}$ being forbidden. Numerical interpolations over the whole range of temperatures (including the normal phase) can be found in Ref.~\cite{pastorechamel2015}.

	\subsubsection{Superfluid phase: subcritical regime $0\leq \Vq<\VLq$}
	
\begin{figure}
		\includegraphics[width=10.5cm]{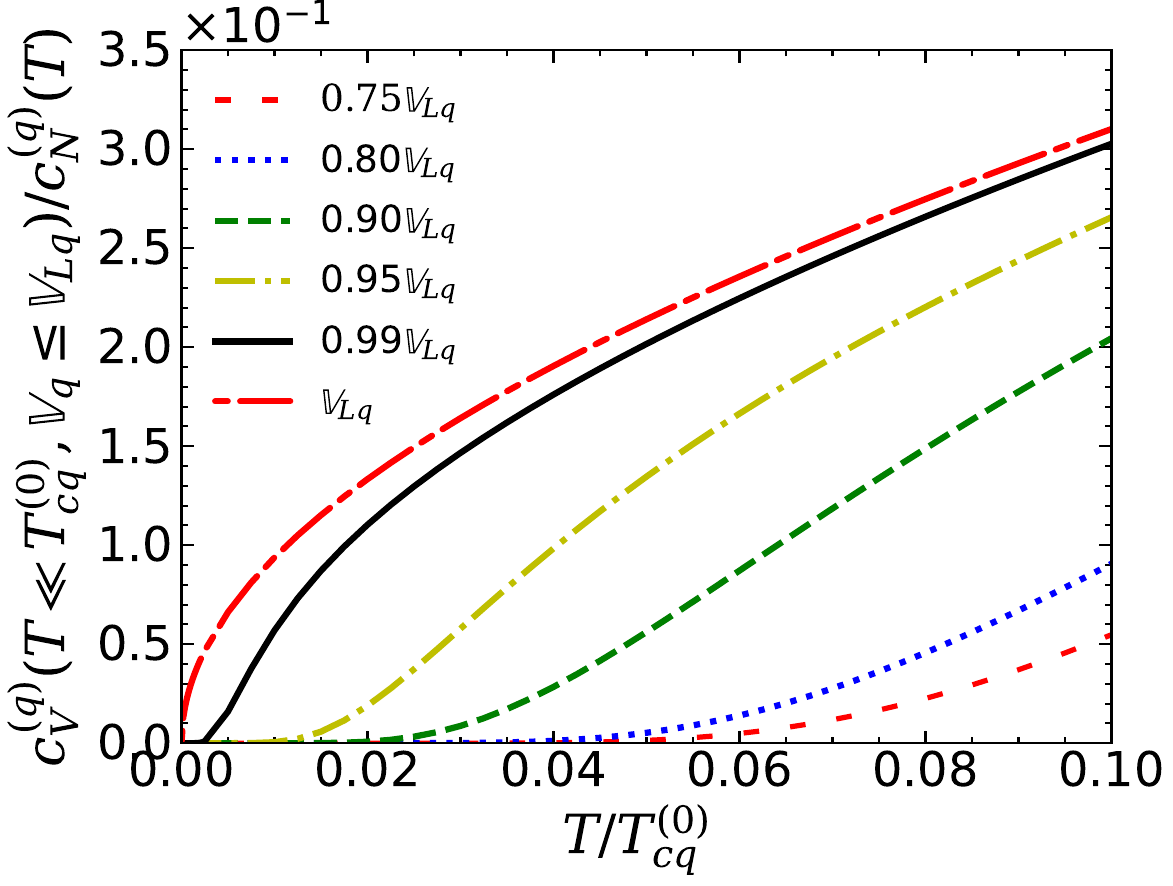}
		\caption{Low-temperature specific heat~\eqref{eq:SpecificHeat-LowT} in subcritical regime (normalized by the specific heat in the normal phase~\eqref{eq:CV-normal}) as a function of the reduced temperature $T/\Tcq$. 
		}
		\label{fig:CvFacteurSubGapless}
	\end{figure}

\begin{figure}
		\includegraphics[width=10.5cm]{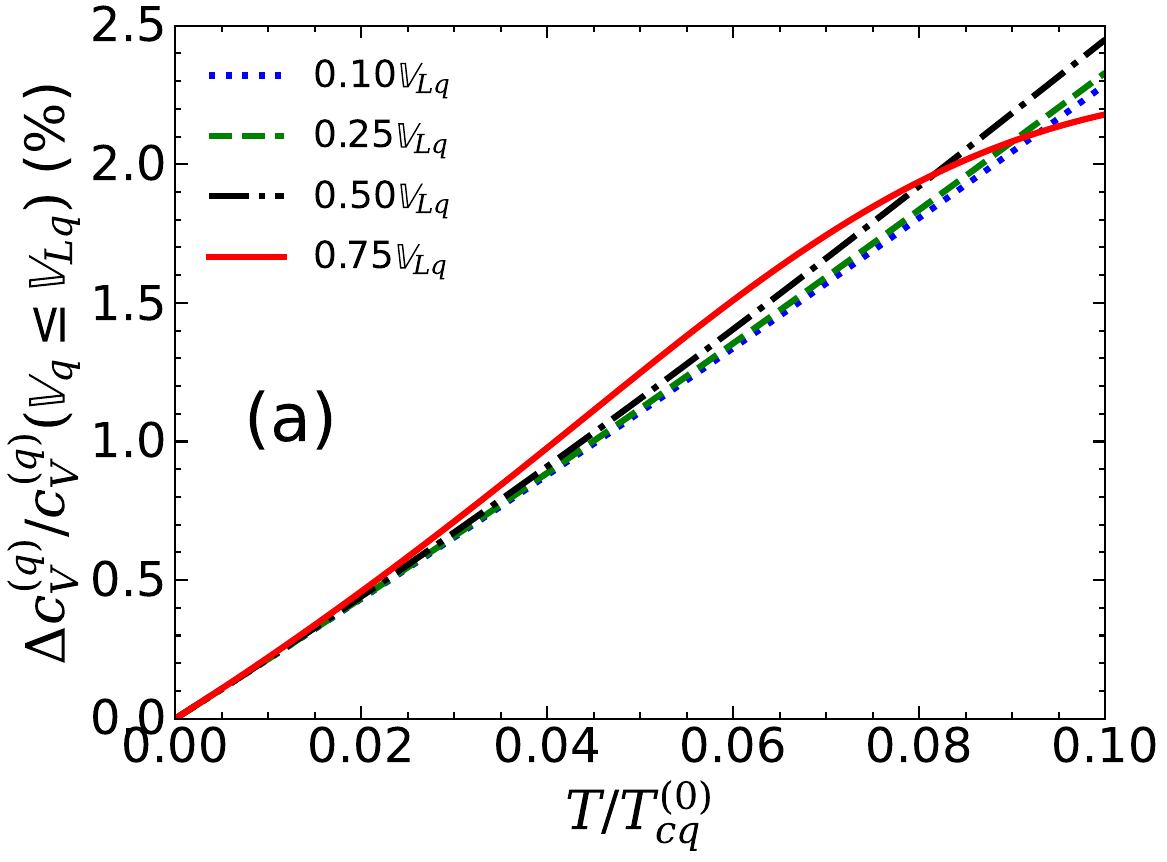}
		\includegraphics[width=10.5cm]{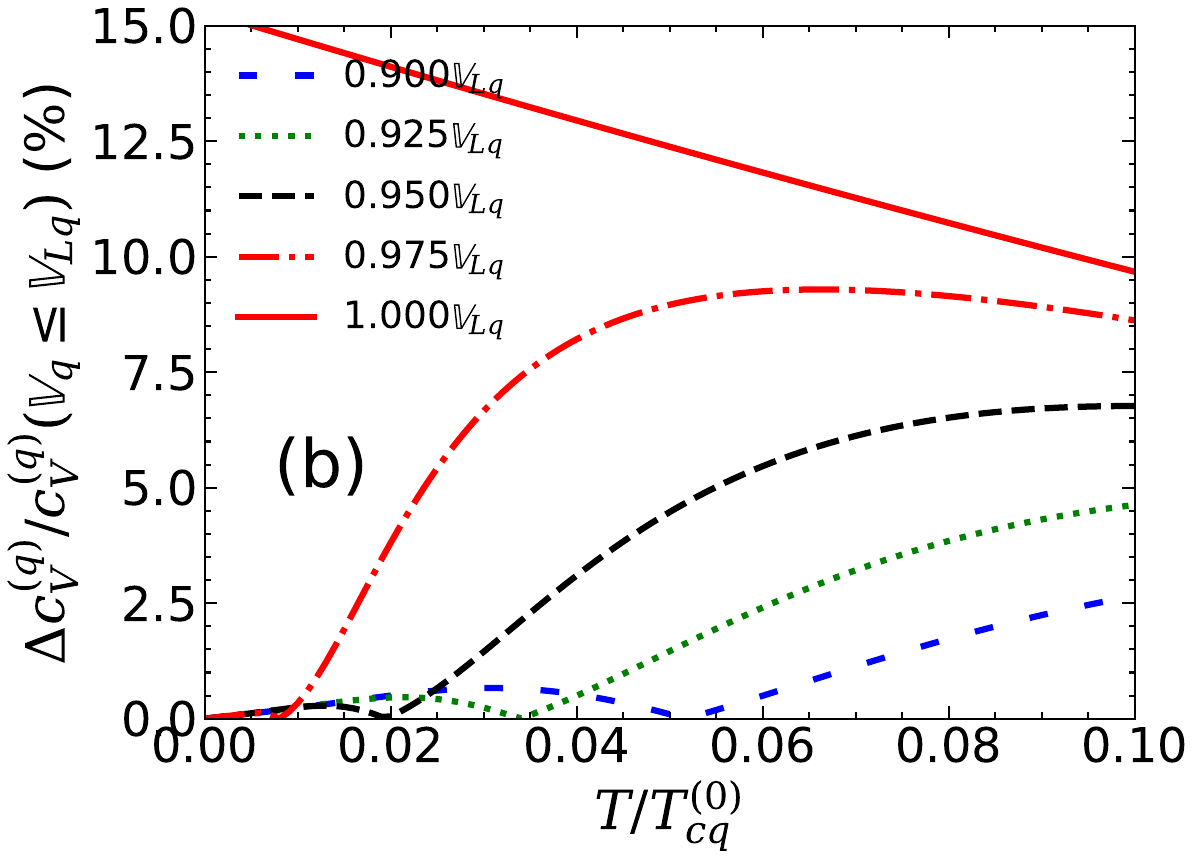}
		
		\caption{Relative deviation (in percents) between the specific heat given  by the ``exact'' expression~\eqref{eq:SpecificHeat-LowT} and the approximation~\eqref{eq:SpecificHeat-SubVLq} as a function of temperature $T/\Tcq$ for different effective superfluid velocities $\Vq$ expressed in terms of Landau's velocity $\VLq$: (a) for effective superfluid velocities $\Vq\leq 0.75 \VLq$, (b) for effective superfluid velocities close to Landau's velocity. The deviation was calculated as (exact-approximate)/exact. } 
		\label{fig:EROverVLSubcritique}
	\end{figure}

	Let us now consider the more general situation of stationary mixtures in presence of superflows in the subcritical regime, i.e. with effective superfluid velocities below Landau's critical velocity. In this case, we notice that $\delta_q^+ > 0$ and $\delta_q^-<0$. For sufficiently low temperatures $T\ll T^{(0)}_{cq}$, we have $\vert\delta_q^\pm\vert \gg 1$ so that both integrals in Eq.~\eqref{eq:SpecificHeat-LowT-Explicit} become vanishingly small. Substituting $\sech^2\left(x/2\right) \approx 4 \exp(-x)$ in Eq.~\eqref{eq:SpecificHeat-LowT-Explicit}, using the approximation $\Delta_q\approx \Delta_q^{(0)}$ (see the discussion in subsection~\ref{subsec:critical-velocities}), and expanding the square roots around $x=\vert\delta_q^\pm\vert$ as   
	\beqy 
	\sqrt{\left(\frac{\expgamma}{\pi}\frac{T}{\Tcq}x\pm \frac{\Vq}{\VLq}\right)^2 -1} \approx \sqrt{2}  \sqrt{\frac{\expgamma}{\pi}\frac{T}{\Tcq}x\pm \frac{\Vq}{\VLq} -1}\, ,
	\eeqy 
	the resulting integrals can then be evaluated analytically. After some simplifications, the specific heat is finally approximately given by 
	\begin{align}\label{eq:SpecificHeat-SubVLq}
		c_V^{(q)}/\Cnorm &\approx \frac{3}{4 \sqrt{2}} \left(\frac{\Tcq}{\expgamma T}\right)^{3/2} \frac{\VLq}{\Vq}\exp\biggl[-\frac{\pi}{\expgamma}\frac{\Tcq}{T}\left(1+\frac{\Vq}{\VLq}\right)\biggr] \\ \notag
		&\times \Biggl\{ \exp\left(\frac{2\pi}{\expgamma}\frac{\Vq}{\VLq}\frac{\Tcq}{T}\right)\left[15 \left(\frac{\expgamma}{\pi}\frac{T}{\Tcq}\right)^2 +\frac{12\expgamma}{\pi}\frac{T}{\Tcq}\left(1-\frac{\Vq}{\VLq}\right)+4\left(1-\frac{\Vq}{\VLq}\right)^2\right] \\ \notag
		&-\left[15 \left(\frac{\expgamma}{\pi}\frac{T}{\Tcq}\right)^2 +\frac{12\expgamma}{\pi}\frac{T}{\Tcq}\left(1+\frac{\Vq}{\VLq}\right)+4\left(1+\frac{\Vq}{\VLq}\right)^2\right]\Biggr\}\, .
	\end{align}
	In the limit $\Vq=0$, it can be shown that this expression reduces to Eq.~\eqref{eq:BCS-SpecificHeat}. For $\Vq>0$, the specific heat is still exponentially suppressed as in the BCS case compared to the normal phase, however the reduction 
	becomes less pronounced with increasing effective superfluid velocity $\Vq$, as can be seen in Fig.~\ref{fig:CvFacteurSubGapless}. The relative deviation between Eq.~\eqref{eq:SpecificHeat-SubVLq} and Eq.~\eqref{eq:SpecificHeat-LowT} does not exceed 2.5\% for effective superfluid velocities $\Vq< 0.9\VLq$ and $T\leq 0.1 T_{cq}^{(0)}$ (see panel (a) of Fig.~\ref{fig:EROverVLSubcritique}) but can reach higher values for velocities closer to Landau's critical velocity (see panel (b) of Fig.~\ref{fig:EROverVLSubcritique}). The low-temperature behavior of the specific heat changes drastically at the onset of the gapless phase. The explicit temperature dependence will be derived in the next subsection.

	\subsubsection{Superfluid phase: onset of gapless regime $\Vq=\VLq$}
	
	The specific heat is amenable to an analytical approximation when the effective superfluid velocity 
	is equal to Landau's critical velocity. 
	Setting $\Vq=\VLq=\Delta_q^{(0)}/(\hbar k_{Fq})$, along with $\Delta_q = \Delta_q^{(0)}$, in Eqs.~\eqref{eq:SpecificHeat-LowT-Explicit} and~\eqref{eq:Bounds} leads to 
	\begin{align}\label{eq:CV-Calculs}
		c_V^{(q)}/\Cnorm&\approx \frac{3}{4\pi^2}\int_0^{+\infty} \text{d}x\; x^2\sech^2\left(\frac{x}{2}\right)\sqrt{\left(\frac{\expgamma}{\pi}\frac{T}{\Tcq}x+1\right)^2-1}\notag\\
		&\qquad - \frac{3}{4\pi^2}\int_{2\pi\Tcq/(T\expgamma)}^{+\infty} \text{d}x\; x^2\sech^2\left(\frac{x}{2}\right)\sqrt{\left(\frac{\expgamma}{\pi}\frac{T}{\Tcq}x-1\right)^2-1} \, . 
	\end{align}
	In the low-temperature limit $T\ll \Tcq$ of interest here, the lower bound of the second integral tends to infinity while the associated integrand vanishes exponentially for high values of $x$ so the contribution of this term to the specific heat can be safely ignored. As for the first integral, we expand the first term inside the square root up to the first order in $\expgamma T x/(\pi \Tcq)$:
	\beqy
	c_{V}^{(q)}/\Cnorm\approx \frac{3}{4\pi^2}\sqrt{\frac{2\expgamma}{\pi}\frac{T}{\Tcq}}\int_0^{+\infty} \text{d}x\; x^{5/2}\sech^2\left(\frac{x}{2}\right)\, .
	\eeqy
	The integral over $x$ can be performed analytically and is given by $\left(4-\sqrt{2}\right)\Gamma\left(7/2\right)\zeta\left(5/2\right)$ ($\Gamma (z)$ and $\zeta(z)$ denotes the Euler gamma function and the Riemann zeta function, respectively). The specific heat in the superfluid phase at Landau's critical velocity is finally given by 
	\begin{align}\label{eq:CV-Landau}
		c_{V}^{(q)}(T\ll \Tcq,\VLq)& = \frac{3}{2\pi^2}\sqrt{\frac{\expgamma}{\pi}}(2\sqrt{2}-1)\Gamma\left(\frac{7}{2}\right)\zeta\left(\frac{5}{2}\right)\sqrt{\frac{T}{\Tcq}}\  \Cnorm(T)\, \notag \\
		&\approx 0.933  \sqrt{\frac{T}{\Tcq}}\  \Cnorm(T)\, .
	\end{align}
	The maximum relative deviations between this approximation and~\eqref{eq:SpecificHeat-LowT} increase linearly with the temperature and do not exceed 5$\%$, for temperatures  $T\leq 0.1 \Tcq$. 
	The low-temperature behavior of the specific heat is very different from the standard BCS case~\eqref{eq:BCS-SpecificHeat} and is only moderately reduced compared to that in the normal phase~\eqref{eq:CV-normal}, as can  be seen in Fig.~\ref{fig:CvFacteurSubGapless}. This is a direct consequence of the gapless superfluidity: the previously 
	forbidden quasiparticle energy gap is now populated thus increasing the available degrees of freedom 
	to store heat. 
	
	\subsubsection{Superfluid phase: gapless regime $\VLq<\Vq \leq \Vcq$}
	
	For effective superfluid velocities larger than $\VLq$, $\Delta_q$ no longer remains constant 
	but decreases with increasing $\Vq$ and vanishes at the critical velocity $\Vcq$. The specific heat~\eqref{eq:SpecificHeat-LowT-Explicit} can be equivalently written as 
	\begin{align}\label{eq:SpecififHeat-LowT-HighV}
		c_V^{(q)}/\Cnorm &\approx \frac{3}{4\pi^2}\int_{-\NegativeBound}^{+\infty} \text{d}x\; \sqrt{\left(\frac{\expgamma}{\pi}\frac{T}{\Tcq}\frac{\VLq}{\Vq}x+1\right)^2-\left(\frac{\Delta_q}{\Delta_q^{(0)}}\frac{\VLq}{\Vq}\right)^2}x^2\sech^2\left(\frac{x}{2}\right)\notag \\
		&\qquad -\frac{3}{4\pi^2}\int_{\PositiveBound}^{+\infty} \text{d}x\; \sqrt{\left(\frac{\expgamma}{\pi}\frac{T}{\Tcq}\frac{\VLq}{\Vq}x-1\right)^2-\left(\frac{\Delta_q}{\Delta_q^{(0)}}\frac{\VLq}{\Vq}\right)^2}x^2\sech^2\left(\frac{x}{2}\right)\, . 
	\end{align}
	For sufficiently low temperatures, the contribution of the second integral can be ignored since the lower bound $\PositiveBound$ in the second integral tends to infinity while its integrand vanishes exponentially. As for the first integral, the square root in the integrand is expanded up to the first order in $\expgamma T\VLq/(\pi \Vq \Tcq)$ 
	\begin{align}\label{eq:expansion-gapless}
		&\sqrt{\left(\frac{\expgamma}{\pi}\frac{T}{\Tcq}\frac{\VLq}{\Vq}x+1\right)^2-\left(\frac{\Delta_q}{\Delta_q^{(0)}}\frac{\VLq}{\Vq}\right)^2} \nonumber \\ 
		&\approx \sqrt{1-\left(\frac{\Delta_q}{\Delta_q^{(0)}}\frac{\VLq}{\Vq}\right)^2} +\frac{\expgamma}{\pi}\frac{\VLq}{\Vq}\frac{T}{\Tcq}x \left[1-\left(\frac{\Delta_q}{\Delta_q^{(0)}}\frac{\VLq}{\Vq}\right)^2\right]^{-1/2}\, .
	\end{align}
	With this, the specific heat reduces to
	\begin{align}
		c_V^{(q)}/\Cnorm &\approx \frac{3}{4\pi^2}\sqrt{1-\left(\frac{\Delta_q}{\Delta_q^{(0)}}\frac{\VLq}{\Vq}\right)^2}\int_{-\NegativeBound}^{+\infty} \text{d}x\; x^2\sech^2\left(\frac{x}{2}\right)\notag \\
		&\qquad +\frac{3\expgamma}{4\pi^3}\frac{\VLq}{\Vq}\frac{T}{\Tcq}\left[1-\left(\frac{\Delta_q}{\Delta_q^{(0)}}\frac{\VLq}{\Vq}\right)^2\right]^{-1/2}\int_{-\NegativeBound}^{+\infty} \text{d}x\; x^3\sech^2\left(\frac{x}{2}\right)\, .
	\end{align}
	Note that this expansion requires $\Vq>\VLq$. 
	Both integrals can be performed analytically and read  
	\beqy
	\int_{-\NegativeBound}^{+\infty}\text{d}x\; x^2 \sech^2\left(\frac{x}{2}\right) &=& 2\biggl\{ \frac{2\pi^2}{3} + (\NegativeBound)^2 \left[\tanh\left(\frac{\NegativeBound}{2}\right)-1\right]  \biggr. \nonumber \\ 
	&&\biggl. + 4\left[\text{Li}_2\left(-\text{e}^{-\NegativeBound}\right) - \NegativeBound\log\left(1+\text{e}^{-\NegativeBound}\right)\right]\biggr\}\, ,
	\eeqy
	\begin{align}
		\int_{-\NegativeBound}^{+\infty}\text{d}x\; x^3 \sech^2\left(\frac{x}{2}\right) &= -24\left[\NegativeBound \text{Li}_2\left(-\text{e}^{-\NegativeBound}\right)+\text{Li}_3 \left(-\text{e}^{-\NegativeBound}\right)\right]\notag \\
		& \qquad -2 (\NegativeBound)^2 \left\{\NegativeBound\left[\tanh\left(\frac{\NegativeBound}{2}\right)-1\right]-6\log\left(1+\text{e}^{-\NegativeBound}\right)\right\}\, ,
	\end{align}
	where we have introduced the the polylogarithm of order $n$ :
	\beqy 
	\text{Li}_n(x)=\frac{x}{\Gamma (n)}\int_{0}^{+\infty} \frac{u^{n-1}}{\text{e}^u - x}\text{d}u\, ,
	\eeqy
	
\begin{figure}
		\includegraphics[width=10.5cm]{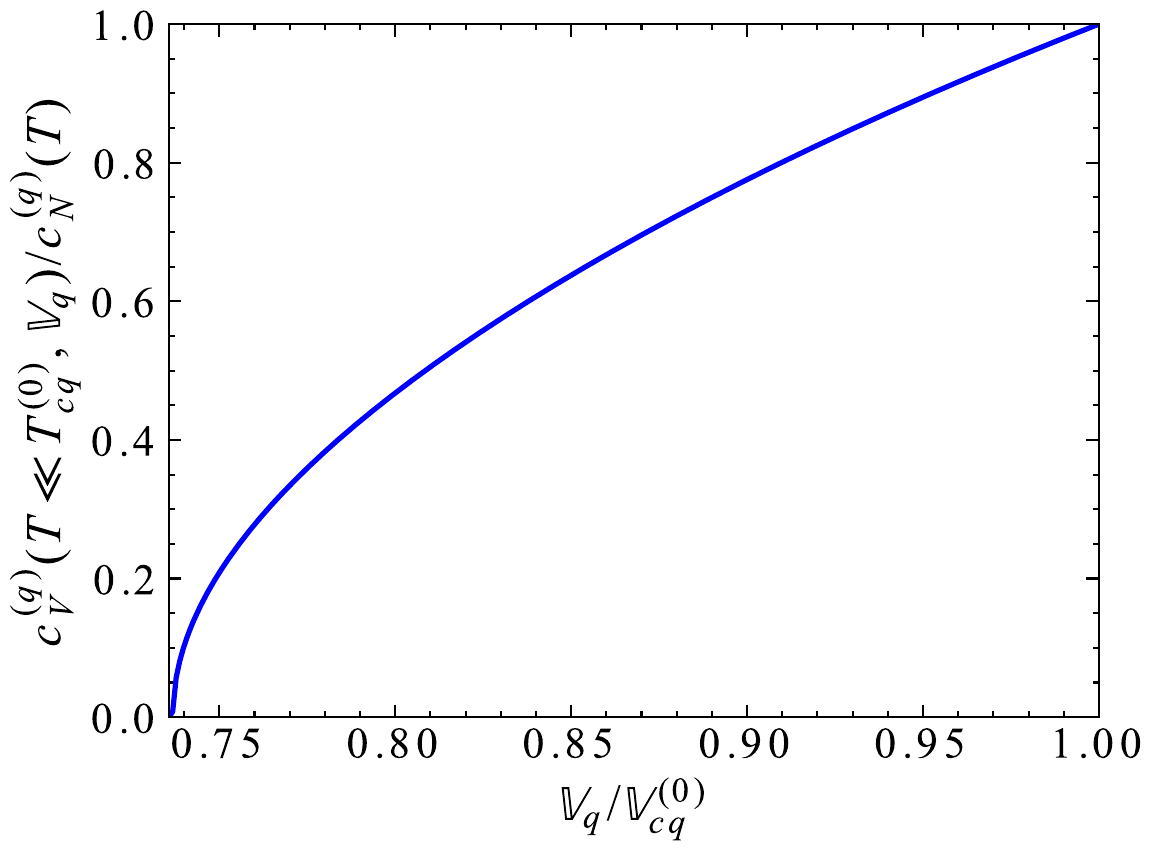}
		\caption{Low-temperature specific heat in gapless regime~\eqref{eq:SpecificHeat-LowT-HighV} (normalized by the specific heat in the normal phase~\eqref{eq:CV-normal}) as a function of the reduced effective superfluid velocity $\Vq/\Vcq$ ranging from $2/\text{e}\simeq0.73576$ (i.e. $\Vq = \VLq$, onset of gapless regime) to 1 (i.e. $\Vq = \Vcq$, disappearance of superfluidity).}
		\label{fig:CvFacteurGapless}
	\end{figure}

\begin{figure}
		\includegraphics[width=10.5cm]{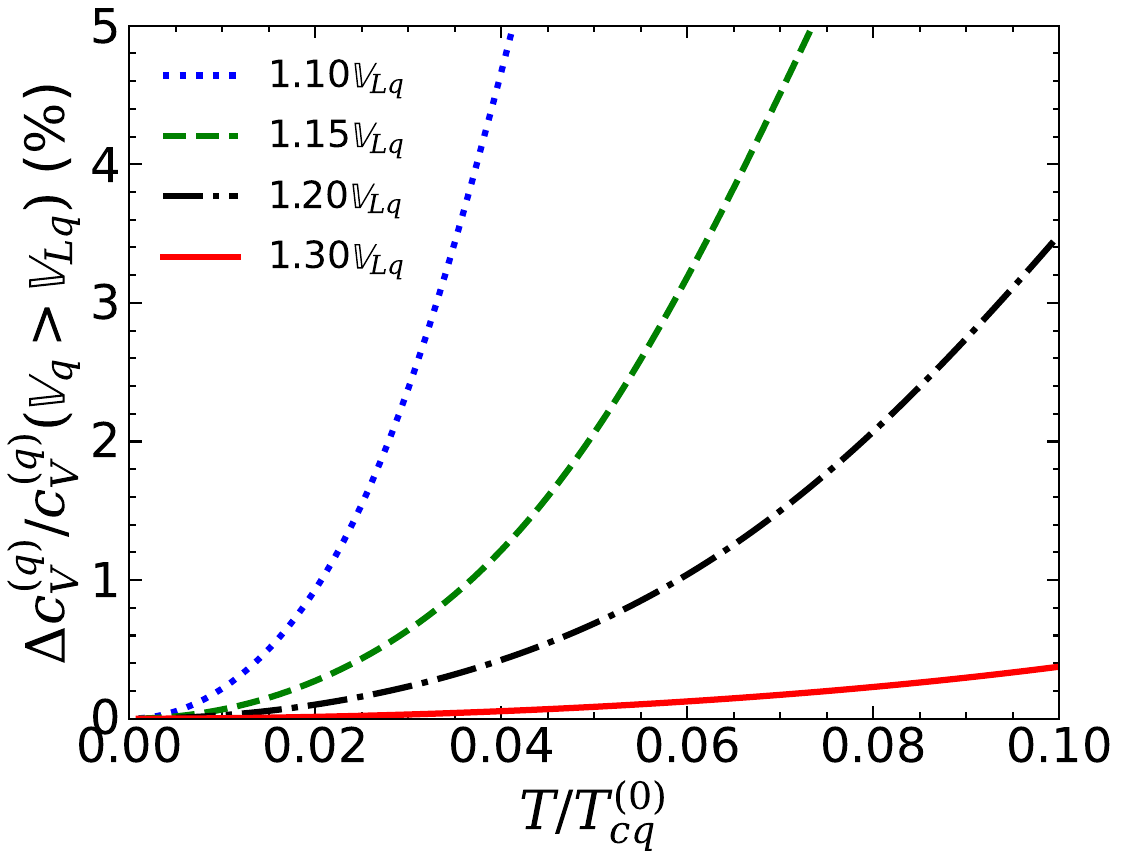}
		\caption{Relative deviation (in percents) between the specific heat given  by the ``exact'' expression~\eqref{eq:SpecificHeat-LowT} and the approximation~\eqref{eq:SpecificHeat-LowT-HighV} as a function of temperature $T/\Tcq$ for different effective superfluid velocities $\Vq$ expressed in terms of Landau's velocity $\VLq$. The deviation was calculated as (exact-approximate)/exact. } 
		\label{fig:EROverVL}
	\end{figure}

	The specific heat is thus given by  
	\beqy
	c_V^{(q)}/\Cnorm &\approx&  \sqrt{1-\left(\frac{\Delta_q}{\Delta_q^{(0)}}\frac{\VLq}{\Vq}\right)^2}\left\{1+\frac{3}{2\pi^2}(\NegativeBound)^2\left[\tanh\left(\frac{\NegativeBound}{2}\right)-1\right]\right. \nonumber \\ &&\left.+\frac{6}{\pi^2}\left[\text{Li}_2(-\text{e}^{-\NegativeBound})-\NegativeBound \log(1+\text{e}^{-\NegativeBound})\right]\right\}\nonumber \\
	&& -\frac{\VLq}{\Vq}\frac{T}{\Tcq}\left[1-\left(\frac{\Delta_q}{\Delta_q^{(0)}}\frac{\VLq}{\Vq}\right)^2\right]^{-1/2}\biggl\{ \frac{18\expgamma}{\pi^3}\left[\NegativeBound \text{Li}_2 (-\text{e}^{-\NegativeBound})+\text{Li}_3 (-\text{e}^{-\NegativeBound})\right]\notag\\
	&&\qquad + \frac{3\expgamma}{2\pi^3} (\NegativeBound)^2 \left[\NegativeBound\left(\tanh\left(\frac{\NegativeBound}{2}\right)-1\right)-6\log (1+\text{e}^{-\NegativeBound})\right]\biggr\}\, .
	\eeqy
	Considering $\NegativeBound\gg 1$ in the hyperbolic tangents and in the (poly)logarithms, the specific heat can finally be approximated by
	\beqy\label{eq:SpecificHeat-LowT-HighV}
	c_V^{(q)}(T\ll \Tcq,\Vq>\VLq)\approx 
	\sqrt{1-\left(\frac{\Delta_q}{\Delta_q^{(0)}}\frac{\VLq}{\Vq}\right)^2} \ \Cnorm(T)\, .
	\eeqy\
	This shows that the specific heat in the gapless regime is comparable to that in the normal phase (see also Fig.~\ref{fig:CvFacteurGapless}) while it is exponentially suppressed in the BCS regime. The specific heat increases 
	with increasing $\Vq$ as more and more quasiparticle states appear in the previously forbidden region. 
	As expected, $c_V^{(q)}$ coincides with the specific heat in the normal phase whenever $\Vq\geq \Vcq$ since then $\Delta_q = 0$ and the superfluidity is destroyed. The relative deviations between the approximation~\eqref{eq:SpecificHeat-LowT-HighV} and Eq.~\eqref{eq:SpecificHeat-LowT} are plotted on Fig.~\ref{fig:EROverVL}.
	For a given $\Vq$, the relative deviation increases with the 
	temperature as expected from our assumption of $T\ll \Tcq$. 
	For a given temperature, the relative deviation increases as the effective superfluid velocity approaches Landau's velocity;  this stems from the fact that the expansion~\eqref{eq:expansion-gapless} becomes singular for $\Vq=\VLq$. At this velocity, Eq.~\eqref{eq:SpecificHeat-LowT-HighV} yields $c_V^{(q)}=0$ independently of the temperature (recalling $\Delta_q \approx \Delta^{(0)}_q$ for $T\ll \Tcq$)  whereas setting  $\Vq=\VLq$ before the expansion leads to the finite value~\eqref{eq:CV-Landau}. 
	
	\section{Conclusions}
	
	Pursuing our previous studies of nuclear superfluidity within the TDHFB theory~\cite{ChamelAllard2019,ChamelAllard2020,ChamelAllard2021}, 
	we have further analyzed the properties of neutron-proton superfluid mixtures.
	
	We have demonstrated that the critical temperatures $\Tcq$ for the disappearance of the neutron and proton superfluid phases in the absence of 
	currents obey the BCS scaling relation~\eqref{eq:Tc0} even though the approximations introduced in the original derivation by Bardeen, Cooper, 
	and Schrieffer in their theory of electron superconductivity~\cite{Bardeen1957} do not hold in the nuclear context. As discussed in 
	our previous studies, the influence of the superflows on the pairing properties are naturally expressed in terms of the effective superfluid 
	velocities $\Vq$ defined by Eq.~\eqref{eq:EffectiveSuperfluidVelocity} rather than the usual superfluid velocities~\eqref{eq:SuperfluidVelocities}. We have 
	found that superfluidity at low temperatures is not destroyed for effective superfluid velocities exceeding Landau's velocity $\VLq$ given by Eq.~\eqref{eq:Landau-velocity-approx} 
	but enters a regime in which the energy spectrum of quasiparticle excitations exhibits no gap while the order parameter $\Psi_q$ therefore also $\Delta_q$ remains finite 
	provided $\Vq$ lies below the critical velocity $\Vcq$ given by Eq.~\eqref{eq:Vcq}. 
	
	We have explicitly studied the vanishing of the quasiparticle energy gap with increasing effective superfluid velocities by calculating 
	the density of quasiparticle states. We have shown that the existence of a gapless superfluid regime has a very strong impact on the specific heat.  
	Whereas $c_V^{(q)}$ is exponentially suppressed in the BCS superfluid phase (compared to the specific heat in the normal phase) at low temperatures and effective superfluid velocities $\Vq<\VLq$, it is only moderately reduced in the gapless regime $\VLq\leq\Vq<\Vcq$. The dependence of $c_V^{(q)}$ on the effective superfluid velocity $\Vq$ is 
	universal when properly rescaled by the critical velocity $\Vcq$ and is very well approximated by Eq.~\eqref{eq:SpecificHeat-SubVLq} in the subcritical regime, and by 
	Eqs.~\eqref{eq:GapInterpolation} and \eqref{eq:SpecificHeat-LowT-HighV} in the gapless phase. 
	The present results may have important implications for the cooling 
	of neutron stars,  as will be discussed in a separate paper~\cite{AllardChamel2023}.

	\begin{acknowledgments}
		This work was financially  supported by the Fonds de la Recherche Scientifique (Belgium) under Grant No. PDR T.004320 and No. IISN 4.4502.19. The authors thank Prof Armen Sedrakian and Dr Wouter Ryssens for discussions.
	\end{acknowledgments}

	\bibliography{references.bib}
	
\end{document}